\documentclass[aps,amsmath,graphicx,twocolumn]{revtex4}

\usepackage{graphics}
\usepackage{graphicx}
\usepackage{amsmath}
\usepackage{amssymb}
\usepackage{epsfig}
\usepackage{epstopdf}
\usepackage{hyperref}

\begin{document}

\title{Two stage pressure-induced Yb valence change in the Hexagonal Laves Phase YbAg$_2$: Investigation by
time differential perturbed angular $\gamma$-$\gamma$ correlation spectroscopy method and density functional calculations}

\author{A.V. Tsvyashchenko}
	\email{tsvyash@hppi.troitsk.ru}
	\affiliation{Vereshchagin Institute for High Pressure Physics, RAS, 108841, Moscow, Troitsk, Russia}
 	\affiliation{Moscow Institute of Physics and Technology, SU, 141700 Dolgoprudny, Russia}
 	
\author{A.N. Nikolaev}
	\affiliation{Skobeltsyn Institute of Nuclear Physics, Moscow State University, 119991 Moscow, Russia}
	\affiliation{Moscow Institute of Physics and Technology, SU, 141700 Dolgoprudny, Russia}
	
\author{D.A. Salamatin}
	\affiliation{Vereshchagin Institute for High Pressure Physics, RAS, 108841, Moscow, Troitsk, Russia}
	\affiliation{Joint Institute for Nuclear Research, 141980, Dubna, Russia}
	\affiliation{Moscow Institute of Physics and Technology, SU, 141700 Dolgoprudny, Russia}
	
\author{A. Velichkov}
	\affiliation{Joint Institute for Nuclear Research, 141980, Dubna, Russia}
    \affiliation{Institute for Nuclear Research and Nuclear Energy, 1784 Sofia, Bulgaria}

\author{A.V. Salamatin}
	\affiliation{Joint Institute for Nuclear Research, 141980, Dubna, Russia}
	
\author{A.P. Novikov}
	\affiliation{Vereshchagin Institute for High Pressure Physics, RAS, 108841, Moscow, Troitsk, Russia}

\author{L.N. Fomicheva}
	\affiliation{Vereshchagin Institute for High Pressure Physics, RAS, 108841, Moscow, Troitsk, Russia}
	
\author{F.S. El'kin}
	\affiliation{Vereshchagin Institute for High Pressure Physics, RAS, 108841, Moscow, Troitsk, Russia}

\author{A.V. Bibikov}
	\affiliation{Skobeltsyn Institute of Nuclear Physics, Moscow State University, 119991 Moscow, Russia}

\author{M.G. Kozin}
	\affiliation{Skobeltsyn Institute of Nuclear Physics, Moscow State University, 119991 Moscow, Russia}
	
\author{M. Budzynski}
	\affiliation{Institute of Physics, M. Curie-Sklodowska University, 20-031 Lublin, Poland}

\begin{abstract}
We have studied the C14 hexagonal Laves phase of YbAg$_2$ at normal conditions and under external pressure up to 19 GPa by the time-differential perturbed angular $\gamma-\gamma$ correlation spectroscopy (TDPAC) using $^{111}$Cd probe nuclei. Under pressure the valence of Yb undergoes a two stage transition from 2.8 to 3. The two stage scenario is characterized by two distinct quadrupole frequencies of $^{111}$Cd probes in silver sublattice, monotonically increasing with pressure and saturating at 8 and 16 GPa. Our experimental data are compared with the density functional studies of the electron band structure of YbAg$_2$, whose results are used for discussion and interpretation of these experiments. We have found that there are two different electric field gradients at inequivalent silver sites and that $4d$-states of silver participate in metal bonding, allowing for the formation of the hexagonal Laves phase.
\end{abstract}
\pacs{PACS here}
\maketitle

\section{Introduction}
\label{sec:introduction}

	Ytterbium intermetallic compounds exhibit a rich variety of interesting physical phenomena such as intermediate valence,  Kondo and heavy fermion behaviors, quantum critical behavior and non-Fermi liquid behavior \cite{1, 2, 3}. Herewith, their properties are highly sensitive to external pressure, magnetic field or temperature as well as to chemical environment of ytterbium ions. This is due to the fact that in the atomic ground state, Yb is divalent with a filled $4f^{14}(sd)^{2}$ shell, but in the solid state, the $f$ electron may experience the electronic transition from the $4f^{14}(sd)^2$ configuration to the $4f^{13}(sd)^3$ configuration (Yb is trivalent). As a consequence, the total $f$ occupancy can be non-integer and fall in between 14 and 13, with the Yb ion valence varying from two to three.
A high valence of ytterbium in elemental metal can be attained by application of pressure \cite{4}.
As pressure mounts from the normal value to 20 GPa, the ytterbium valence increases from two to an intermediate state ($2.55 \pm 0.05$, Ref.\ \cite{5}) reaching the value of three above 100 GPa \cite{6}. Similarly to the pristine metal,
in intermetallic compounds the Yb ion can exhibit different integral valence (two or three) or be in the intermediate
valence regime \cite{7,8,9,10,11,12,13,14}.

Changeable ytterbium valence is also observed in novel metastable compounds synthesised under high pressure and temperature conditions \cite{15,16,17,18}.
However, despite much experimental effort, the trivalent state of Yb in these intermetallic compounds has not been achieved even at sufficiently high pressures \cite{17,19,20,21}.
Here we report on our pressure studies of the novel hexagonal phase of the YbAg$_2$ compound,
demonstrating a two-stage evolution of Yb valence,
which terminates at trivalent electronic state.
Investigations of the effect of pressure on the valence of ytterbium in YbAg$_2$ are of particular interest since it is probably the best
way to detect subtle changes in metallic bonding.
Earlier we have shown that the effect of high pressure can be studied within the time-differential perturbed angular $\gamma-\gamma$ correlation spectroscopy (TDPAC) using $^{111}$Cd probe nuclei \cite{22}, and established a linear correlation between the valence and the electric field gradient
(or the quadrupole frequency $\nu_Q$).
Here we present
a pressure-induced investigation where we have used diamond anvil cell to generate pressure up to 19 GPa.

At ambient pressure YbAg$_2$ has the orthorhombic structure of the CeCu$_2$ type \cite{23}, which is non-magnetic with the divalent state of Yb.
Since the divalent state of ytterbium implies conventional behaviour, for our experiments we have chosen
the hexagonal phase of YbAg$_2$ crystallized in the (C14) MgZn$_2$ structure (at pressure of 5 GPa),
where the valence of ytterbium is 2.8 \cite{18}.
In the hexagonal phase a silver atom can be easily substituted with the $^{111}$Cd probe and as a result we can create
a relatively high density of $^{111}$Cd nuclei required for the TDPAC measurements of hyperfine interactions parameters under high pressure.

The details of TDPAC measurements are given in Sec.\ \ref{sec:methods}, the obtained results are discussed in Sec.\ \ref{sec:results}.
For interpretation of the results in Sec.\ \ref{sec:results}
we have used {\it ab initio} calculations of the electronic band structure of YbAg$_2$ and the related compound where some silver atoms
are substituted with Cd-probes.

\section{Experiment and theory}
\label{sec:methods}

\begin{figure}
\resizebox{0.2\textwidth}{!}
{
\includegraphics{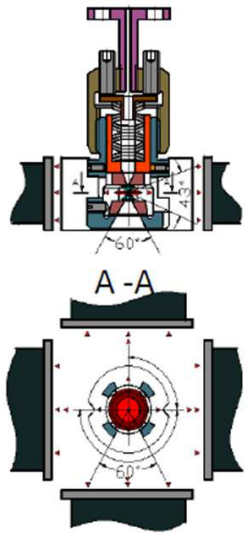}
}
\caption{Diamond anvil cell with four surrounding windows for four BaF$_2$ detectors.
Lower panel represents the A-A cross-section.}
\label{fig:1}
\end{figure}

	Experimental measurements were carried out by the TDPAC method using the 171 -- 245 keV $\gamma-\gamma$ cascade in $^{111}$Cd populated through the 2.8 d isotope $^{111}$In electron capture decay. The cascade proceeds via the 245 keV level with the half-life $\tau_{1/2} = 84$ ns, spin $I = 5/2$, and quadrupole moment $Q$ = 0.83(13) b \cite{Herzog_Q}. The $^{111}$In activity was produced via the $^{109}$Ag ($\alpha, 2n$) $^{111}$In reaction through irradiating a silver foil with the 32 MeV $\alpha$-beam at the Nuclear Physics Institute cyclotron (Moscow State University). The $^{111}$In - $^{111}$Cd nuclear probes were introduced into the lattice of YbAg$_2$ by the high-pressure synthesis: the constituents (Yb and Ag) taken in proper amounts with an overall weight of about 600 mg were melted together with a  piece of the irradiated silver foil (1 mg) in a special chamber under pressure of 5 GPa \cite{18}. The X-ray diffraction pattern presented in Fig.\ \ref{fig:1} confirms the hexagonal Laves phase \cite{18}.
	As the samples were polycrystalline and paramagnetic at temperatures
above 4 K, the perturbation of the angular correlation can be described by the perturbation factor $G_{22}$ for the static electric quadrupole interaction (EQI) \cite{27}:

\begin{eqnarray}
G_{22} (t; \nu_Q, \eta, \sigma) =  \nonumber \\
\sum_{i}^{} p_i (S_{20} +
\sum_{n}^{} S_{2n} cos(\omega_nt)exp(-\omega_n^2\sigma^2t^2/2)) .
\label{eq:g22}
\end{eqnarray}

	Here $p_i$ stands for the relative populations of nonequivalent sites of the probe nuclei; the hyperfine frequencies $\omega_n$ depend on the quadrupole coupling constant $\nu_Q=eQV_{zz}/h$ called below the quadrupole frequency (QF) and the asymmetry parameter $\eta = (V_{xx} - V_{yy})/V_{zz}$, where
$V_{ii} = \frac{\partial^2 V}{\partial^2 i} (i = x, y, z)$ are the principal-axis components of the tensor of the electric field gradient (EFG). The coefficients $S_{2n}$ depend only on $\eta\ (0 \leq \eta \leq 1)$.  For the nuclear spin $I = 5/2$ there are three transitions: $n = 1, 2, 3$. The exponential term is of the gaussian form customarily used to represent a possible distribution of EFG and $\sigma$ is the relative half-width of the Gaussian distribution. Here we restrict ourselves to the perturbation parameter of the second order since the unperturbed angular correlation coefficient $A_{44} \ll A_{22}$ ($A_{22} = -0.18$).
	The perturbation factor $G_{22}(t)$ describing a nuclear spin precession due to the hyperfine interaction, was determined in a usual way from the TDPAC spectrum $R(t)$, measured by a 4-detector spectrometer \cite{26}.
The TDPAC spectrum $R(t)$ was obtained by combining the delayed coincidence spectra measured at the angles of $\pi/2$ and $\pi$ between detectors, $N(\pi/2, t)$ and $N(\pi, t)$, through the expression

\begin{eqnarray}
\label{eq:aniz_theory}
R(t) = -A_{22} Q_2 G_{22}(t)
\end{eqnarray}

Here $Q_2 \approx 0.80$ is the solid-angle correction and

\begin{eqnarray}
\label{eq:aniz_exp}
R(t) = -2\frac{N(\pi, t) - N(\pi/2, t)}{N(\pi, t) + 2N(\pi/2, t)}
\end{eqnarray}

\begin{figure}
\resizebox{0.45\textwidth}{!}
{
\includegraphics{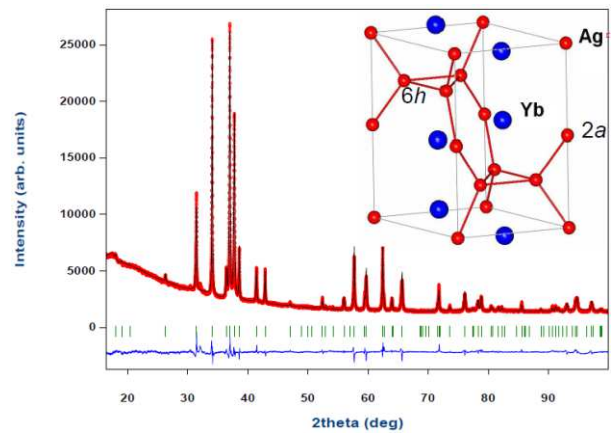}
}
\caption{X-ray powder diffraction profile (red) and its Rietveld refinement fit (black) for the hexagonal (C14) sample of YbAg$_2$ with $^{111}$Cd probes.
The difference plot (blue) and Bragg peaks (green) are plotted underneath. Inset shows the C14 hexagonal structure of YbAg$_2$.}
\label{fig:2}
\end{figure}

	External pressures up to 8 GPa were generated in a calibrated `toroid'-type device \cite{28}.
For pressures above 8 GPa we used a special high-pressure cell with diamond anvils with $\approx 580$ $\mu$m culet diameters and $\approx \pi/3$ aperture.
The diamond anvil cell is shown schematically in Fig. \ref{fig:2}.
A 200 $\mu$m hole was drilled in the pre-indented stainless--steel gasket. Samples of YbAg$_2$ ($\approx 75 \times 75 \times 25$ $\mu$m$^3$ in size), doped with $^{111}$In-$^{111}$Cd and ruby chips were  placed inside the gasket filled with sodium chloride as a pressure-transmitting medium. Pressure was measured by the ruby luminescence technique using the standard ruby calibration scale.


The quadrupole frequencies $\nu_{Q}(6h)$ (the $^{111}$Cd localized in $6h$ sites) and $\nu_{Q}(2a)$ (the $^{111}$Cd localized in $2a$ sites), where $\nu_Q = eQV_{zz}/h$ were measured as a function of pressure (up to 19 GPa) at room temperature and as a function of temperature at normal pressure.

Experimental data were analyzed with the DEPACK software \cite{DEPACK}.

The electronic band structure of YbAg$_2$ has been calculated using the linearized augmented plane wave method (LAPW) with the potential of general shape taking into account a semicore 5$p$ band of Yb and the spin-orbit coupling splitting the 4$f$ and 5$p$ Yb band electron shells \cite{LAPW-M}.
The C14 primitive unit cell is large: in total there are 12 atoms of three different types 4Yb, 2Ag($2a$), 6Ag($6h$)
(a detailed discussion on two different silver sites is given in Sec.\ \ref{sec:results}).
With the LAPW basis state
cut off parameter $R^{max}_{MT}$ $K_{max}$ = 9 we have 1131 basis functions for each point of the Brillouin zone.
The number of $k-$points
was 215 for the self-consistent procedure and 1331 for the final run, with the number of calculated low-lying bands about 100.
For calculation of the exchange-correlation potential and the exchange-correlation energy contribution we have used a variant of
the local density approximation
within the density functional theory (DFT) \cite {LDA}.
Calculations with Cd probe atom have been carried out by substituting Ag with Cd in the site $2a$ or $6h$ of the YbAg$_2$ unit cell.

\section{Results and discussion}
\label{sec:results}

   The x-ray diffraction measurements were carried out on a set of YbAg$_2$ powdered samples with the $^{111}$Cd probes, from which the C14
hexagonal structure (the $P6_3/mmc$ space symmetry, the group number 194, $Z = 4$) illustrated in Fig.\ \ref{fig:1} was confirmed \cite{18}.
The diffraction pattern and optimized structure obtained from Rietveld refinement (Fig. \ref{fig:1})
yields the lattice constants $a = b = 5.6853(2)$ {\AA} and $c = 9.3105(4)$ {\AA}.
%
\begin{table}
\caption{ Nearest neighbor shells and distances (in {\AA}) with the number of atoms in each shell for three nonequivalent atoms of YbAg$_2$.
Last line shows lowest spherical harmonics used in the density expansion around each element.
\label{tab1f} }

\begin{ruledtabular}
\begin{tabular}{c  c  c  c  }

        shell & Yb  & Ag($2a$) & Ag($6h$) \\
\tableline
            1 & 3.32812 6Ag($6h$) &  2.85482 6Ag($6h$)  & 2.82435 2Ag($6h$)  \\
            2 & 3.33396 3Ag($2a$) &  3.33396 6Yb        & 2.85482 2Ag($2a$)  \\
            3 & 3.34977 3Ag($6h$) &  4.65998 2Ag($2a$)  & 2.85718 2Ag($6h$)  \\
            4 & 3.46777 1Yb       &  4.91901 12Ag($6h$) & 3.32812 4Yb  \\
            5 & 3.49017 3Yb       &  5.22254 6Yb        & 3.34977 2Yb  \\
\tableline
      $Y_L^m$ & $L = 1$, 2, 3, 4, ... & $L = 2$, 4, ... & $L = 1$, 2, 3, 4, ... \\

\end{tabular}
\end{ruledtabular}
\end{table}

\begin{figure}
\resizebox{0.38\textwidth}{!}
{
\includegraphics{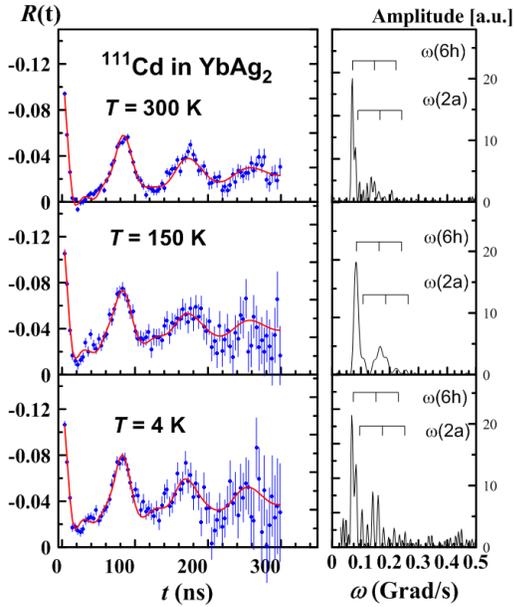}
}
\caption{(left panels) Temperature evolution of the TDPAC spectrum $R(t)$ of YbAg$_2$ taken from the $^{111}$Cd nuclear probes located at
crystallographically nonequivalent $2a$ and $6h$ sites of the Ag sublattice at normal pressure. (right panels) Their Fourier transforms. }
\label{fig:3}
\end{figure}

\begin{figure}
\resizebox{0.38\textwidth}{!}
{
\includegraphics{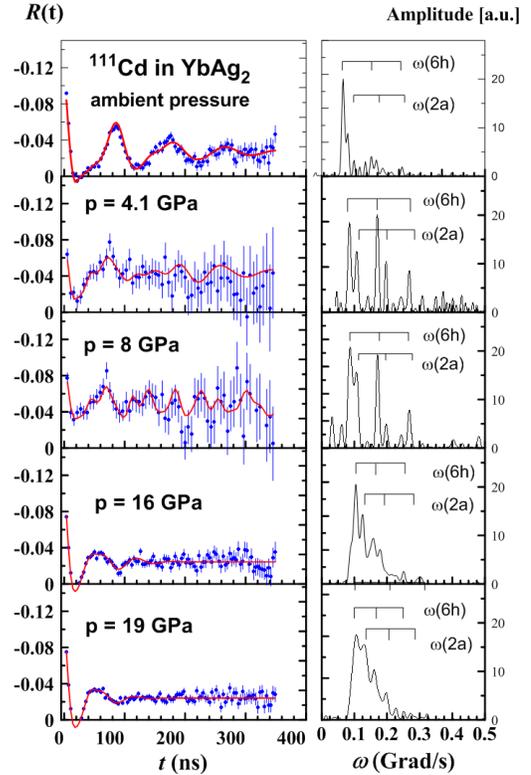}
}
\caption{(left panels) Pressure evolution of the TDPAC spectrum $R(t)$ of YbAg$_2$ taken from the $^{111}$Cd nuclear probes located at
crystallographically nonequivalent $2a$ and $6h$ sites of the Ag sublattice at room temperature. (right panels) Their Fourier transforms.
The spectra at 4.1 and 8 GPa have large errors because the measurements were carried out in a toroid-type chamber, for which a low signal-to-background ratio was observed.}
\label{fig:4}
\end{figure}

In the C14 phase there are two different sites of silver, Ag($2a$) and Ag($6h$), with the population ratio $p(2a)/p(6h) = 1 : 3$
(see Fig. \ref{fig:1} and Table \ref{tab1f}).
The Ag($2a$) atom in the $2a$ position is surrounded by 6 Ag($6h$) and 6 Yb atoms.
The Ag($6h$) atom in the $6h$ position is surrounded by 4 Ag($6h$) and 2 Ag($2a$) atoms, and 6 Yb atoms.
(Detailed information on neighboring shells for all atoms is given in Table \ref{tab1f}.)
Ag($2a$) and Ag($6h$) sites have different local symmetry. The Ag($2a$) site has the inversion symmetry while the Ag($6h$) site does not.
As a result, the electron density around the Ag($2a$) site is expanded in spherical harmonics $Y_L^m$ with even $L$ (i.e., $L = 0$, 2, 4, ...),
while the electron density around  Ag($6h$) expands in spherical harmonics with even and odd $L$ (i.e., $L = 0$, 1, 2, 3, ...).
Both silver sites (Ag($2a$) and Ag($6h$)) allow for a quadrupole electric component ($l = 2$) and a nonzero electric field gradient (EFG),
which enables the TDPAC signal. However, since the crystallographic sites and density expansion of the two silver sites are different,
EFG for Ag($2a$) and Ag($6h$) are expected to be different too. This fact is confirmed by the electron band structure calculation of YbAg$_2$.
In the TDPAC experiments we also observed nonequivalent EFGs on cadmium probe atoms inserted at sites $2a$ and $6h$.
The TDPAC spectra $R(t)$ measured at the $^{111}$Cd probe nuclei in YbAg$_2$ are given in Figs.\ \ref{fig:3}, \ref{fig:4},
whose panels reproduce the evolutions of $R(t)$ with temperature at normal pressure and with pressure at room temperature.

It is worth noting that the $^{111}$Cd probe nuclei occupy exclusively sites of the Ag sublattice.
This has been demonstrated previously for the YMn$_2$ compound synthesized in the same hexagonal structure at high pressure \cite{29}.
Another indication of this conclusion is that the experimentally observed occupation ratio $p(6h)/p(2a)$ is close to three,
clearly corresponding to the  ratio of $6h/2a$ sites in the Ag sublattice
(in the unit cell of the C14 structure there are 2 Ag($2a$) atoms and 6 Ag($6h$) atoms).
Since there are two types of Ag sites [Ag($2a$) and Ag($6h$)] having different values of EF gradients,
there are two quadrupole frequencies  $\nu_{Q}(2a)$ and $\nu_{Q}(6h)$
associated with the $^{111}$Cd probe nuclei located at these sites.
These frequencies as well as other characteristics of EFGs are quoted in Table~\ref{tab:table1}, \ref{tab:table2}.

\begin{table*}[b]
\caption{
Hyperfine quadrupole interaction parameters measured at the $^{111}$Cd probe nuclei in YbAg$_2$ at various pressures at room temperature.
$\nu_{Q}(6h)$, $\eta(6h)$, $\sigma(6h)$, $p(6h)$ and $\nu_{Q}(2a)$, $\eta(2a)$, $\sigma(2a)$, $p(2a)$ stand for the quadupolar frequency,
the asymmetry parameter, the frequency distribution and the relative population of the $^{111}$Cd probes located at crystallographically nonequivalent
$6h$ and $2a$ sites of the Ag sublattice, respectively.
}
\label{tab:table1}
\begin{ruledtabular}
\begin{tabular}{c|c|c|c|c|c|c|c|c}
  $P$ (GPa), $T = 300$ K & $\nu_Q(6h)$ (MHz)& $\nu_Q(2a)$ (MHz)& $\eta(6h)$ & $\eta(2a)$ & $\sigma(6h)$ & $\sigma(2a)$ & $p(6h)$ & $p(2a)$ \\ \hline
  0 & 76(1) & 97(3) & 0.2(1) & 0.0(1) & 0.1(1) & 0.1(1) & 0.8(1) & 0.3(1) \\ \hline
  4.1 & 80(2) & 98(2) & 0.4(1) & 0.4(1) & 0.2(1) & 0.2(1) & 0.8(1) & 0.2(1) \\ \hline
  8.0 & 93(2) & 107(2) & 0.4(1) & 0.3(1) & 0.05(2) & 0.05(2) & 0.5(1) & 0.3(1) \\ \hline
  16.0 & 93(3) & 123(4) & 0.5(1) & 0.5(1) & 0.15(5) & 0.1(1) & 0.7(1) & 0.3(1) \\ \hline
  19.0 & 91(3) & 125(5) & 0.4(1) & 0.5(1) & 0.18(4) & 0.18(4) & 0.7(1) & 0.3(1) \\
\end{tabular}
\end{ruledtabular}
\end{table*}

\begin{table}
\caption{
Measured ($V^{exp}_{zz}$) and calculated ($V^{calc}_{zz}$) EFGs in 10$^{21}$ V/m$^2$ at $P = $ 0 and 19 GPa at room temperature.
}
\label{tab:table2}
\begin{ruledtabular}
\begin{tabular}{ c|c|c|c|c|c|c }
  YbAg$_2$ (C14) & \multicolumn{2}{c|}{$V^{exp}_{zz}$ at $^{111}$Cd} & \multicolumn{2}{c|}{$V^{calc}_{zz}$ at Cd} & \multicolumn{2}{c}{$V^{calc}_{zz}$ at Ag} \\ \hline
  $P$, $T$ & $6h$ & $2a$ & $6h$ & $2a$ & $6h$ & $2a$ \\ \hline
  0 GPa, 300 K & 3.8(6) & 5.0(8) & 1.9 & 4.1 & 6.1 & 8.0 \\ \hline
  19 GPa, 300 K & 4.5(7) & 6.0(10) & 2.0 & 4.7 & 6.9 & 9.1 \\
\end{tabular}
\end{ruledtabular}
\end{table}
	
    From Fig. \ref{fig:3} showing the temperature evolution of the TDPAC data,
it follows that the frequencies $\nu_{Q1}$ and $\nu_{Q2}$ practically do not change with temperature.
Consequently, in the region 4--300 K the valence of ytterbium is approximately temperature independent.

\begin{figure}[]
\resizebox{0.45\textwidth}{!}
{
\includegraphics{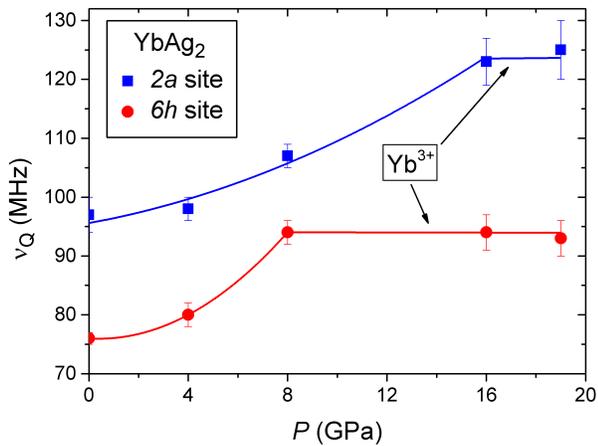}
}
\caption{Pressure evolution of the quadrupole frequencies $\nu_Q$ for $^{111}$Cd nuclear probes located at
crystallographically nonequivalent sites ($2a$ and $6h$) of Ag in YbAg$_2$.}
\label{fig:5}
\end{figure}

    Another important finding is a variation of both quadrupole frequencies with pressure.
The $6h$ site frequency, $\nu_{Q}(6h)$, is increased from 76 to 93 MHz (at $P=8$ GPa), while the $2a$ site frequency, $\nu_{Q}(2a)$,
changes from 100 to 125 MHz (at $P=16$ GPa).
These changes of $\nu_{Q}(6h)$ and $\nu_{Q}(2a)$ with pressure are summarized in Fig.\ \ref{fig:5}.
From Fig.\ \ref{fig:5} we see that a characteristic feature of both plots is a smooth rise to a certain value (93 MHz at 8 GPa for $\nu_{Q}(6h)$,
125 MHz at 16 GPa for $\nu_{Q}(2a)$), which then remains constant with further pressure increase.
The constant value of $\nu_{Q}(2a,6h)$ is an indication that Yb in YbAg$_2$ has reached a stable monovalent state,
which in our case is the trivalent one.
Indeed, earlier it has been shown that GdAl$_2$ with the trivalent state of Gd shows no significant change of the quadrupole frequency \cite{22}.
	
In a mixed valent compound the pressure ($P$) or valence ($\upsilon = \upsilon(P)$) dependence of the measured quadrupole frequency $\nu_Q = \nu_Q(\upsilon)$
is commonly described by the linear equation $\nu(\upsilon)= \nu_2 + (\nu_3 - \nu_2) (\upsilon - 2)$.
In our case $\upsilon$ is the valence of Yb (between two and three),
while $\nu_2$ ($\nu_3$) is the frequency measured for the compound with the divalent (trivalent) ytterbium.
In YbAg$_2$ at normal pressure $\upsilon_0 = 2.8$ \cite{18}, and for the divalent state of Yb we accept $\nu_2 = 5.7$~MHz.
(This frequency can be extracted from data for CaAl$_2$.)
The linear relation then can be rewritten in respect to $\nu_3$, $\nu_3 = \nu_2 + (\nu(\upsilon_0) - \nu_2) /(\upsilon_0 - 2)$.
Substituting here the experimental ($P=0$) quadrupole frequency values $\nu(\upsilon_0) = 97$ MHz for the $2a$ site and 76 MHz for the $6h$ site
we obtain
that $\nu_3 = 125$ MHz and 93 MHz for the $2a$ and $6h$ site, respectively.
These values are in good correspondence with the saturated
frequencies $\nu_{Q}(2a) \approx 125$ MHz and $\nu_{Q}(6h) \approx 93$ MHz
shown in Fig.\ \ref{fig:5}, indicating self-consistency of the obtained experimental results demonstrating the trivalent state of ytterbium.

The existence of two distinct frequencies, $\nu_{Q}(2a)$ and $\nu_{Q}(6h)$, in YbAg$_2$ at normal and elevated pressure, Table~\ref{tab:table1}
and Fig.\ \ref{fig:5}, and their different pressure dependence lead to the conclusion that the electronic properties of silver at
two crystallographically nonequivalent sites ($6h$ and $2a$) are fairly different and
silver atoms located at the $6h$ and $2a$ sites behave to some extent as different elements.

\begin{figure}[b]
\resizebox{0.45\textwidth}{!}
{
\includegraphics{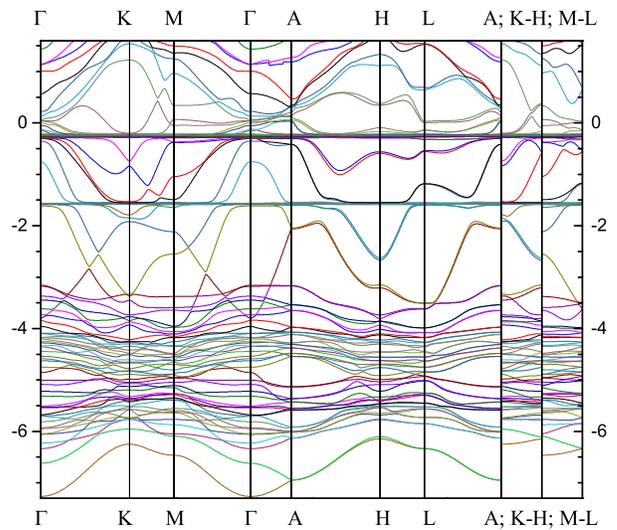}
}
\caption{Electron band structure of YbAg$_2$ along high symmetry lines. The Fermi level ($E_F$) corresponds to the zero energy.
Calculated semicore 5$p$ band of Yb lie about 27 eV below $E_F$ (not shown).}
\label{fig:6}
\end{figure}

The conclusion is fully confirmed by {\it ab initio} band structure calculations, whose main results are presented in Figs.\ \ref{fig:6}.
First, one can notice two very narrow group of band states centered around -1.57 eV and -0.24 eV, which correspond to splitted 4$f_{5/2}$ and 4$f_{7/2}$
electron levels. The region from -7 eV to -3 eV refers to the silver 4$d$ states hybridized with other ytterbium states (mainly with 5$d$ and 6$s$).
Notice that the 4$d$ states are not
separated from the 5$s$ and other Yb states by a gap of forbidden states, and therefore the 4$d$ states
fully participate in the formation of metal bonding in YbAg$_2$.

\begin{figure}[t]
\resizebox{0.36\textwidth}{!}
{
\includegraphics{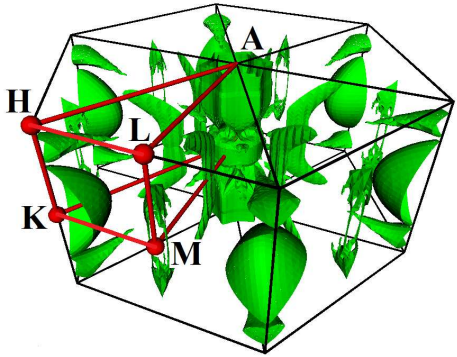}
}
\caption{The Fermi surface of YbAg$_2$.}
\label{fig:7}
\end{figure}

Our calculations have revealed a complex shape of the Fermi surface of YbAg$_2$, shown in Fig.\ \ref{fig:7}.
Its complexity reflects a relatively large
intersections of the Fermi energy by valence electron bands. The calculated density of states (DOS) at the Fermi level
is 3.86 state/eV or 9.09 mJ/mol K$^2$. The last value should be compared with 12.9 mJ/mol K$^2$ found from the experimental fit of
specific heat \cite{18}. The discrepancy is usually accounted for by the electron-phonon interaction.
In the present case there can be also an influence of the low temperature anomaly developed below 6 K (Fig.\ 5 of Ref.\ \cite{18}).

\begin{figure}[t]
\resizebox{0.45\textwidth}{!}
{
\includegraphics{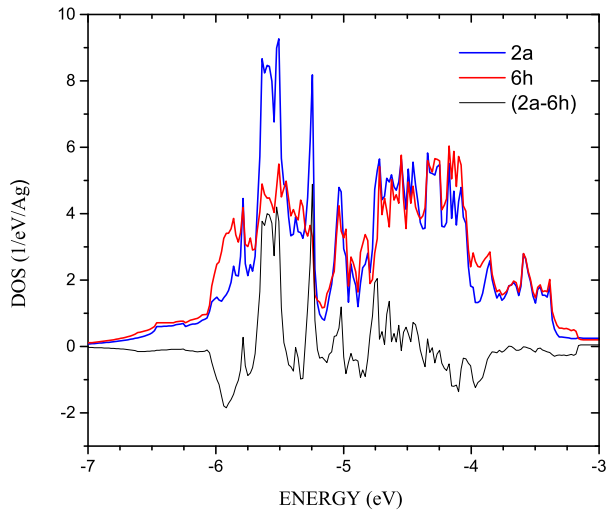}
}
\caption{Density of states (DOS) of two different silver atoms [Ag($2a$) and Ag($6h$), see text for details] in YbAg$_2$.
Black line stands for the difference plot.}
\label{fig:8}
\end{figure}

Of particular interest is the
partial density of states (DOS) of the two silver atoms and their difference shown in Fig.\ \ref{fig:8}.
Notice, that DOS of Ag($2a$) is larger than DOS of Ag($6h$) at the energy range from -5.45 eV to -5.68 eV and also around -5.24 eV and -4.74 eV,
whereas at other energies the opposite takes place.
Partial charges ($Q_l$, where $l=s$, $p$, $d$, $f$) of two silver atoms are close but nevertheless different. A typical difference is 0.01.
At the $2a$ site the values of the partial charges $Q_l$ with even $l$ (i.e. $Q_s$, $Q_d$) are somewhat larger while at the $6h$ site $Q_l$ with odd $l$
(i.e. $Q_p$, $Q_f$) are larger. The total electron charge is slightly larger for Ag($2a$).
The difference between Ag($2a$) and Ag($6h$) is much more pronounced for multipole density distributions and charges, that is,
for density components with $L \neq 0$.
For example, the quadrupole electron density component at $2a$ site is described by the spherical component $Y_{l=2}^0$, whereas for
$6h$ site it is given by two angular functions, $S_1 = -c_1 Y_{l=2}^0 + c_2 Y_{l=2}^{2,c}$ and $S_2=c_2 Y_{l=2}^0 + c_1 Y_{l=2}^{2,c}$.
Here $c_1=0.7559$, $c_2=0.6547$ and $Y_{l=2}^{2,c}$ stands for the real spherical harminics of the $cos-$type.
Different quadrupole distribution of electron charges results in different EFG at silver nuclei.
For EFG at Ag($2a$) we have -8.04 10$^{21}$ V/m$^2$ and for EFG at Ag($6h$) -6.12 10$^{21}$ V/m$^2$.
Calculations with Cd yield -4.11 V/m$^2$ for Cd at the $2a$ site and -1.90 10$^{21}$ V/m$^2$ ($\eta = 0.1$) for Cd at the $6h$ site.
(It should be noted that the calculation of Cd in the $6h$ site is less precise than in the $2a$ site. Such a calculation is only approximate
because during the calculation we ignore a symmetry change induced by the probe Cd-atom. The symmetry of the $2a$ site on the other hand
 does not change with substitution of Ag by Cd).
Therefore, EFG is larger at the $2a$ site in the correspondence with the experimental data.
Since the electron density at the $6h$ site is described by two functions $S_1$ and $S_2$ with the azimuthal component $m=2$, the matrix of
gradients is not axially symmetric, and the EFG asymmetry parameter $\eta \neq 0$. This fact is also in correspondence with the EFG data
measured at $^{111}$Cd in $6h$ sites at normal conditions, Table \ref{tab:table1}.
We have $\eta = 0.8$ for Ag($6h$) and $\eta = 0.1$ for Cd($6h$).
We have also calculated that the minimum of total energy of YbAg$_2$ associated with the trivalent state of Yb around the 6$h$-site of Ag and found that it corresponds to the unit cell
volume $V$(Yb$^{+3}$, $6h$-site) = 245.3 {\AA}$^3$. This volume is approximately 6{\%}
smaller than the experimental one at normal pressure and our estimations show that it can be reached under pressure of $\approx8.5$~GPa.
Notice that this pressure leads to the saturated value of the $\nu_Q$ at the $6h$-site marking the onset of the Yb trivalent state, Fig.\ \ref{fig:5}.
Calculations also confirm the experimental increase of EFG with pressure, Fig.\ \ref{fig:5} and Table \ref{tab:table2}.
The calculated values of EFG with $V$(Yb$^{+3}$, $6h$-site) are -9.07 10$^{21}$ V/m$^2$ at Ag($2a$), -6.94 10$^{21}$ V/m$^2$ with $\eta = 0.8$ at Ag($6h$) for YbAg$_2$;
-4.69 10$^{21}$ V/m$^2$ at Cd($2a$), -2.0 10$^{21}$ V/m$^2$ with $\eta= 0.46$ at Cd($6h$) for Cd probes in YbAg$_2$.
It should be mentioned that a slight increase of the asymmetry parameter $\eta$ with pressure is also observed for the $2a$ site.
This may be connected with deformation of the crystal lattice due to the non-hydrostatic properties of NaCl as pressure-transmitting medium.

The valence of ytterbium is found to be close to three at normal pressure. Formally, the number of 4$f$ states $Q_f = $ 13.68 per Yb atom.
However, if following Ref.\ \cite{Stra} we localize 13 $f$ electrons during the self-consistent procedure, then we obtain that the residual number of 4$f$ band states is close to null. On the other hand, for elemental ytterbium which is in the divalent state, the number of 4$f$ band states is very close to 14.
These findings according to arguments of Ref.\ \cite{Stra} implies that the ytterbium has an intermediate valence close to three.

\section{Conclusions}

Our TDPAC study confirms that the valence of ytterbium in the hexagonal C14 phase of YbAg$_2$ is intermediate (2.8 \cite{18}) and close to three.
The valence of Yb in the hexagonal phase is different from the divalent state of Yb found in the orthorhombic phase of the CeCu$_2$ structure.
Under external pressure we observe an unusual behaviour of quarupole frequencies shown in Fig.\ \ref{fig:5} which we characterize as a two stage transition
to the trivalent state of Yb. A saturation of the TDPAC quadrupole frequency is first occurred for probes at the $6h$ sites at 8 GPa ($\nu_{Q}(6h)$) and
then for probes at the $2a$ sites ($\nu_{Q}(2a)$) at a still higher pressure of 16 GPa.
We attribute the two stage transition to two different states of silver at the $2a$ and $6h$ sites of the C14 structure
and correspondingly to two different states of the $^{111}$Cd probes.
A thorough discussion on this issue supported by DFT band structure calculations of YbAg$_2$
is presented in Sec.\ \ref{sec:results}.
The $4d$-states of silver are involved in metal bonding, making silver in YbAg$_2$ analogous to other $4d$ transition metals (for example, Ru) with
unoccupied $4d$-shell and allowing for the formation of the hexagonal Laves phase.
It is also remarkable that the transition to the trivalent state takes place in YbAg$_2$ under a relatively moderate pressure of 8-16 GPa.

\acknowledgments

The authors are grateful to S.~M. Stishov, N.~G. Chechenin and V.~B. Brudanin  for support of this work.
The work is supported by the Russian Foundation for Basic Research (Grants No. 16-02-01122, and No. 17-02-00064)
and by special program of the Department of Physical Science, Russian  Academy of Sciences.
The work at the Joint Institute for Nuclear Research was carried out under the auspices of a
Polish representative in the JINR. A.V.T. and D.A.S. (TDPAC measurements) acknowledge the support
from Russian Scientific Foundation (Grant RNF 17-02-01050).
The research is carried out using the equipment of the shared research facilities of HPC
computing resources at Lomonosov Moscow State University


\end{document}